# Diode-like Selective Enhancement of Carrier Transport through Metal-Semiconductor Interface Decorated by Monolayer Boron Nitride

Hemendra Nath Jaiswal,[1] Maomao Liu,[1] Simran Shahi,[1] Sichen Wei,[2] Jihea Lee,[2] Anindita Chakravarty,[1] Yutong Guo,[1] Ruiqiang Wang,[1] Jung Mu Lee,[1] Chaoran Chang,[2] Yu Fu,[2] Ripudaman Dixit,[1] Xiaochi Liu,[3] Cheng Yang,[4] Fei Yao,[2,*] and Huamin Li[1,*]

[1] *Department of Electrical Engineering, University at Buffalo, The State University of New York, Buffalo, New York 14260, US*

[2] *Department of Materials Design and Innovation, University at Buffalo, The State University of New York, Buffalo, New York 14260, US*

[3] *School of Physics and Electronics, Central South University, Changsha 410083, China*

[4] *School of Physics and Electronics, Shandong Normal University, Jinan 250014, China*

[*] Author to whom correspondence should be addressed. Electronic address: feiyao@buffalo.edu and huaminli@buffalo.edu

# Abstract

Two-dimensional (2D) semiconductors such as monolayer molybdenum disulfide ($MoS_2$) are promising material candidates for next-generation nanoelectronics. However, there are fundamental challenges related to their metal-semiconductor (MS) contacts which limit the performance potential for practical device applications. In this work, we exploit 2D monolayer hexagonal boron nitride (*h*-BN) as an ultrathin decorating layer to form a metal-insulator-semiconductor (MIS) contact, and design an innovative device architecture as a platform to reveal a novel diode-like selective enhancement of the carrier transport through the MIS contact. The contact resistance is significantly reduced when the electrons transport from the semiconductor to the metal, but barely affected when the electrons transport oppositely. We propose a concept of carrier collection barrier to interpret this intriguing phenomenon as well as a negative Schottky barrier height obtained from temperature-dependent measurements, and show the critical role of the collection barrier at the drain end for the overall transistor performance.

# Introduction

Two-dimensional (2D) monolayer molybdenum disulfide (MoS$_2$) as an n-type semiconducting transition metal dichalcogenides (TMDs) has been demonstrated as a promising material candidate for future energy-efficient nanoelectronics [1-4] because of ultrathin body with enhanced electrostatic gating, natural carrier confinement, suitable bandgaps, passivated surfaces, good intrinsic carrier mobility, and mechanical flexibility, etc [5-11]. To fully explore the potential of monolayer MoS$_2$ for practical device and circuit applications, there is a critical need of metal-semiconductor (MS) contact engineering and optimization which can lead to the maximization of device performance [12-14]. Especially for the extremely-scaled short-channel devices, the contact condition between 2D semiconductors and three-dimensional (3D) metals plays a more important role in the entire carrier transport process [15, 16]. To lower Schottky barrier height (SBH) at the MS interface and thus reduce contact resistance ($R_c$), a variety of approaches for MoS$_2$-based nanoelectronic devices have been proposed. On the 2D MoS$_2$ side, substitutional doping [13], surface charge transfer doping [17-20], isoelectronic alloying [21-23], hybridization [24], and phase engineering [25-26] can increase the carrier density of MoS$_2$ and thus reduce the contact barrier height. On the 3D metal side, although the effect of the conventional work function engineering following the Schottky-Mott rule is limited primarily due to the unique Fermi level pinning effect in MoS$_2$ [27-29], it has been demonstrated that Al [30], Sc and Ti [31] contacts can provide the low SBHs for electrons, compared to MoO$_x$ [32], graphene oxide (GO) [33], and NbS$_2$ [34] with the low SBHs for holes. Novel contact architecture such as one-dimensional (1D) edge contact [35, 36] and high-quality metal deposition condition such as ultra-high vacuum [37] can also yield a low $R_c$ value.

In addition to the engineering on 2D MoS$_2$ and 3D metals, contact decoration by introducing ultrathin foreign materials at the interface has been proposed as a new strategy to improve the MS contact. For example, graphene as an electrically active semimetal can form a 2D-2D MS contact with MoS$_2$, and the gate-controlled Fermi level tunability of graphene can assist to reduce the SBHs [38-41]. As a comparison, the metal-insulator-semiconductor (MIS) contact has been proposed as well. Various ultrathin insulators such as Ta$_2$O$_5$ [42], TiO$_2$ [43-44], and MgO [45] have been demonstrated to reduce $R_c$, introduce the Fermi level de-pinning effect, and consequently improve the device performance, primarily due to the reduction of the SBH, the suppression of metal-induced gap states and the formation of electric dipoles. As an atom-thick 2D layered insulator, hexagonal boron-nitride (*h*-BN) with a bandgap of ~6 eV has been exploited in the MIS contact to de-pin the Fermi level and lower $R_c$ [46-50]. In principle, the insulating decoration layer in the MIS contact configuration should be thick enough to suppress the MS interfacial interaction, yet thin enough to provide a high tunneling probability for carrier transport through the insulator. As a result, there is a trade-off between the dominations of barrier height and tunneling resistance to obtain the lowest $R_c$.

In this work, we designed an innovative platform where both the MS and MIS contacts can be achieved and compared on a single monolayer MoS$_2$ triangle domain. First, we demonstrated the improved contact condition and consequently the boosted device performance of MoS$_2$ FETs by using the MIS contact decorated by a monolayer *h*-BN. Then, we discovered a novel diode-like selective enhancement of the carrier transport through the MIS contact, where the MIS contact can significantly reduce $R_c$ and augment the electron transport from the semiconductor to the metal, but has negligible effects on the electron transport oppositely. Finally, we exploited a concept of carrier collection barrier in a comparison with the conventional carrier injection barrier to understand the selective enhancement of the carrier transport through the MIS contact as well as a negative SBH obtained from temperature-dependent measurements, and showed the critical role of the collection barrier at the drain end for the overall transistor performance.

## Results

**Device structures.** For a comparative study, we fabricate three types of global-back-gate transistor architectures and measure four different transistor configurations depending on the assignment of source and drain: an MS-source MS-drain (MS-MS) FET, an MIS-source MIS-drain (MIS-MIS) FET, an MS-source MIS-drain (MS-MIS) FET, and an MIS-source MS-drain (MIS-MS) FET, as shown in **Fig. 1** and **Fig. S1** in **Supporting Information**. The MS contact is made by a Ti/Au (10 nm/100 nm) layer and a monolayer MoS$_2$ (~0.65 nm), and the MIS contact is made by adding a monolayer $h$-BN (~0.4 nm) [51, 52] between the Ti/Au layer and the monolayer MoS$_2$. Both MoS$_2$ and $h$-BN are synthesized by chemical vapor deposition (CVD) [53, 54], and their monolayer structures are confirmed by Raman spectroscopy and atomic force microscopy (AFM). Especially, we design and fabricate all three types of the device architectures on a monolayer MoS$_2$ triangular domain with identical geometries including channel length and width (see **Methods**). Therefore, all the devices share the same quality of the MoS$_2$ channel, and the difference of the device performance can be solely attributed to their contact conditions. We also fabricate the devices for transmission line measurement (TLM) to extract $R_c$ with the MS and MIS contacts.

**Output characteristics.** For all four transistor configurations, a comparison of drain current density ($J_D$) versus drain voltage ($V_D$) at various gate voltages ($V_G$) is performed in both linear and logarithmic scales at room temperature, as shown in **Fig. 2**. There are several features to note. First, the $J_D$-$V_D$ characteristics are not symmetric for the positive and negative $V_D$ even with the same type of the contacts at the source and drain, due to an asymmetric contact condition (i.e., the different contact areas and thus the different $R_c$ at the source and drain). Since each device is fabricated on a triangular corner, the device has a trapezoidal channel. We define the shorter metal contact as the source and the longer one as the drain for both the MS-MS and MIS-MIS FETs. The asymmetric factor, defined as the ratio of the current density magnitude at the positive $V_D$ to that at the negative $V_D$, can be calculated as a function of $V_D$ for the MS-MS and MIS-MIS FETs, as shown in **Fig. S2** in **Supporting Information**. Second, the MS-MS FET at $V_G = 40$ V shows the highest $J_D$ of 1.2 nA/µm, and it increases up to 19.3 nA/µm for the MIS-MIS FET. For both the MS-MIS and MIS-MS FETs, the highest magnitude of $J_D$ has an intermediate value of about 7 nA/µm. The results clearly indicate that the introduction of the MIS contact, either at the source or drain end, can enhance the carrier transport and increase the current density. Third, a Schottky contact is observed in all the FETs which have at least one MS contact (i.e., the MS-MS, MS-MIS, and MIS-MS FETs), yet a quasi-Ohmic contact is obtained in the MIS-MIS FET. The superior performance of the MIS contact, including both the increased current density and the Schottky-to-Ohmic contact transition, can be interpreted by the quantum tunneling and barrier lowering which are induced by the decoration of the monolayer $h$-BN at the MS interface. Specifically, the conventional MS contact between Ti and $MoS_2$ forms the Schottky barrier with an interacted interface, and the surface potential, defined as the difference between the metal and semiconductor work function, completely locates at the semiconductor side. Whereas the MIS contact has a monolayer $h$-BN and a van der Waals (vdW) gap (~0.3 nm) [55] between Ti and $MoS_2$. These additional barriers with a total thickness of ~1 nm is still thin enough to allow the quantum tunneling occur with a high tunneling probability, but is thick enough to share the original surface potential, so the band bending on the $MoS_2$ side and thus the effective SBH are reduced, giving rise to a smaller $R_c$.

To further investigate the underlying mechanism of the MIS contact, here we include the asymmetric factor at $V_G = 40$ V and reconstruct the output characteristics at the same condition, as shown in **Fig. 3 (a-c)**. Here the $V_D$-dependent asymmetric factor is an average based on the values obtained from both the MS-MS and MIS-MIS FETs (see **Fig. S2** in **Supporting Information**), and is applied to eliminate the asymmetry between the source and drain contacts. Compared to the MS-MS FET, it is intriguing to see that the MIS contact, either at the drain in the MS-MIS FET or at the source in the MIS-MS FET, shows a novel and strong selectivity for the carrier transport through it. The electron transport from the semiconductor to the metal is significantly enhanced by a factor of ~7 (e.g., at $V_D = 0.5$ V for the MIS drain contact in **Fig. 3 (a)** and at $V_D = -0.5$ V for the MIS source contact in **Fig. 3 (b)**), but the transport from the metal to the semiconductor is negligibly affected (e.g., at $V_D = -0.5$ V for the MIS drain contact in **Fig. 3 (a)** and at $V_D = 0.5$ V for the MIS source contact in **Fig. 3 (b)**). Whereas, the MIS-MIS FET does not possess this selectivity anymore, due to a back-to-back connection of two MIS contacts which are equivalently identical for the carrier transport from either direction (see **Fig. 3 (c)**). The maximum $J_D$ obtained at $V_G = 40$ V in this FET is enhanced by a factor of ~ 20 which is higher than those in any other device.

Considering the description of $J_D$-$V_D$ characteristics at a specific $V_G$ condition as $J_D = V_D/R_{tot} = V_D/(R_{cS}+R_{cD}+R_{ch})$, it is possible to visualize the impact of the MIS contact on the $V_D$-dependent resistances, as shown in **Fig. 3 (d-f)**. Here $R_{tot}$, $R_{cS}$ (or $R_{cD}$), and $R_{ch}$ are the total resistance, the contact resistance at the source (or at the drain), and the channel resistance, respectively. It is clear to see that the value of $R_{tot}$ is significantly reduced by introducing the MIS drain contact at the positive $V_D$ or the MIS source contact at the negative $V_D$. Because both the MS $R_c$ and $R_{ch}$ are consistent for all the devices in comparison, the reduction of $R_{tot}$ directly indicates a decrease of $R_c$ ($\Delta R_c$) by exploiting the MIS contact, as shown in **Fig. 3 (g)**. The value of $\Delta R_C$ can reach up to ~400 Ωm (for the MIS drain contact at $V_D = 0.5$ V and for the MIS source contact at $V_D = -0.5$ V) only when the electrons transport from the semiconductor to the metal, but approximates zero for the electron transport oppositely. Furthermore, by taking the sum of $\Delta R_c$ from the MIS-MS and MS-MIS FETs, the overall reduction of $R_c$ for the MIS-MIS FET can be estimated as a function of $V_D$, as shown in **Fig. 3 (h)**. The calculated $\Delta R_C$ in this method is in a good agreement with the experimental results which is extracted from the difference of $R_{tot}$ between the MS-MS and MIS-MIS FETs. Its symmetric behavior for both the positive and negative $V_D$ is also consistent with the experimental data, serving as good evidence for the proposed theory.

We further investigate the dependence of the carrier transport on $V_D$ by plotting $\ln(J_D/V_D^2)$ versus $1/V_D$ curves, as shown in **Fig. 4**. For the MS-MS FET, no linear dependence is found even $V_D$ is up to 2 V. However, by introducing the MIS drain contact in the MS-MIS FET, *J-V* linearity induced by Fowler-Nordheim Tunneling (FNT) appears when $V_D > 0.12$ V (see **Fig. 4 (a)**). Considering the identical MS source contact and the channel condition, it is clear that the drain contact plays an extremely important role in the overall transistor performance. The electron transport along the channel is governed by both the injection barrier at the source and the collection barrier at the drain, and thus the applied $V_D$ is divided by $R_{cS}$, $R_{ch}$, and $R_{cD}$. Due to the significant reduction of $R_{cD}$ by introducing the MIS drain contact, the effective $V_D$ drop over $R_{cS}$ at the MS source contact becomes much higher compared to that in the MS-MS FET, and thus easily reaches to the threshold to enable the FNT at the MS source contact. In contrast, for the case of the MIS-MIS FET, a more effective $V_D$ drop is allocated over $R_{ch}$ due to the reduction of both $R_{cS}$ and $R_{cD}$.

Therefore, the carrier injection at the source is still dominated by direct tunneling (DT) and thermionic emission (TE) (see **Fig. 4 (b)**). The FNT at the source cannot occur unless a higher $V_D$ is applied.

**Transfer characteristics.** The $J_D$-$V_G$ transfer characteristics of all the types of the FETs are shown in **Fig. 5 (a)**. The MIS-MIS FET shows the best performance in terms of on-current density, on-off ratio, and subthreshold swing, whereas the MS-MS FET has the worst performance. The MS-MIS and MIS-MS FETs show the intermediate performance, and their difference is attributed to the asymmetric channel geometry and the assignment of the source and drain contacts. Similar results, together with the Schottky-to-Ohmic contact improvement, have also been reproduced from other devices using a thinner insulator for the back gating, as shown in **Fig. S3** in **Supporting Information**.

For all the types of the FETs, their maximum $J_D$ at $V_D = 0.5$ V is obtained at $V_G = 40$ V, and their temperature ($T$) dependence is measured from 203 to 303 K, as shown in **Fig. 5 (b)** and **Fig. S4** in **Supporting Information**. As $T$ increases, all the devices show an increase of $J_D$ except for the MS-MIS FET. The virtual independence of the temperature suggests the quantum mechanical tunneling as the predominant carrier transport mechanism in the MS-MIS FET, which is also consistent with the demonstration of the FNT in our prior discussion (see **Fig. 4 (a)**). As a comparison, the MIS-MS FET shows a similar behavior like the MS-MS FET, because the MIS source contact under the positive $V_D$ barely affects the electron transport from the metal to the semiconductor (see **Fig. 3 (b)** and **(e)**). The field-effect mobility ($\mu_{FE}$), defined as $(L/W)(1/C_{ox})(1/V_D)(\partial I_D/\partial V_G)$, is calculated from the transfer characteristics, and the maximum values as a function of $T$ are shown in **Fig. 5 (c)**. Here $L$ is the channel length, $W$ is the average channel width, and $C_{ox}$ is the capacitance of 285-nm-thick SiO$_2$. It has to be mentioned that the direct comparison of the $\mu_{FE}$ values is not appropriate due to the asymmetry in the source and drain contacts as well as the channel geometry. However, their dependence on $T$, described by a power law as $\mu_{FE} \sim T^{\gamma}$, can directly indicate the difference in the carrier transport mechanism. It is intriguing to see that all the devices with the MIS contact at the drain, either in the MIS-MIS or MS-MIS FETs, possess a strong and consistent temperature dependence of $\mu_{FE}$, and the exponent $\gamma$ is obtained as ~0.6. As a comparison, for all the devices with the MS contact at the drain, either in the MS-MS or MIS-MS FETs, the value of $\mu_{FE}$ has a relatively weak temperature dependence and $\gamma$ approximates to ~0.1. This result clearly indicates that the drain contact for carrier collection plays an important role in the overall carrier transport and transistor performance. This result is also consistent with our prior discussion on the diode-like selective enhancement: the MIS contact can significantly affect the electron transport from the semiconductor to the metal (e.g., the MIS contact at the drain under the positive $V_D$) but barely vary the transport on the opposite direction (e.g., the MIS contact at the source under the positive $V_D$).

## Discussion

Following the discovery of the diode-like selective enhancement of the carrier transport through the MIS contact, we further investigate its impact on $R_c$ and SBH using the TLM and $T$-dependent measurements. Two TLM devices are fabricated with an identical structure, but one has the MS contacts and another one has the MIS contacts. The value of $R_c$ is extracted from a linear predication at room temperature and plotted as a function of $V_G$, as shown in **Fig. 6 (a)**, where the off-state, on-state, and subthreshold regions are identified based the transfer characteristics in **Fig. 5 (a)**. In both the off-state and on-state regions, $R_c$ with the MS contact is higher compared to that with the MIS contact, but becomes lower in the subthreshold region.

The values of the SBH are extracted from the *T*-dependent output and transfer characteristics [18, 56, 57] in the MS-MS and MIS-MIS FETs, and they illustrate several important features, as shown in **Fig. 6 (b)** and **Fig. S5** in **Supporting Information**. First, for both the MS-MS and MIS-MIS FETs, the SBH varies from positive to negative as the devices switch from the off state to the on state. The negative SBH has also been reported in the *h*-BN-decorated [48] or graphene-decorated metal contacts [58] not only for 2D materials but also for the conventional semiconductors [59]. Various underlying mechanisms are proposed and they are still under debate, for example, the overshadow effect by high series resistance [59, 60] and the electric-field-driven modulation effect of the work functions [58, 61, 62]. Based on our experimental data, here we propose a new theory using the concept of the collection barrier at the drain for interpretation. In the conventional approach to extract the thermionic SBH, the positive SBH is used to describe the energy potential barrier which prevents the electron transport from the metal to the semiconductor at the source, and the contact at the drain is assumed to be an ideal Ohmic contact. However, if the drain contact is not an ideal case, the barrier at the drain could also prevent the electron collection, and thus the negative SBH can appeal which describes the barrier preventing the carrier collection from the semiconductor to the metal. Here we take the MIS-MIS FET as an example, and the energy band diagram with an applied positive $V_D$ [63] is shown in **Fig. 6 (c)**. When the transistor is at the off state, the electron transport is primarily limited by the injection barrier at the source, and the drain contact acts as an Ohmic contact. As $V_G$ increases, the transistor moves into the subthreshold region. When the transistor is at the on state, the electron transport is mainly limited by the collection barrier at the drain, and the source contact is virtually transparent due to a high tunneling probability. Even the injected electrons have a relatively higher energy, most of the energy would be dissipated by collisions during the transport along the channel, and eventually face the collection barrier at the drain with the relatively low energy. Second, the absolute SBH value of the MIS-MIS FET, either for the carrier injection (positive SBH) or carrier collection (negative SBH), is always lower in the off- and on-state regions than that in the MS-MS FET but higher in the subthreshold region (see **Fig. 6 (b)**). This result is very consistent with the $V_G$-dependent $R_c$ obtained from the TLM devices (see **Fig. 6 (a)**), serving as good evidence to prove the concept of the collection barrier. Third, although the collection barrier increases with $V_G$, $J_D$ still increases. This is because the 2D carrier density in the channel increases significantly with $V_G$. Considering the actual collection barrier height obtained in this work is only ~30 meV at the maximum, its effect on the current is completely masked by the enlarged 2D carrier density as $V_G$ increases.

Moreover, based on the energy band diagrams of the collection barrier (see **Fig. 6 (c)**), one can expect that the carrier transport mechanism through the MIS drain contact is a combination TE and quantum tunneling. When the collection barrier dominates, the electrons with high kinetic energies transport from the semiconductor to the metal by the DT, and the ones with low energies transport by the TE and thermionic field emission (TFE). As $T$ increases, the electron distribution in the conduction band is widened to a higher energy range, and thus more electrons can tunnel through the MIS drain contact. This result is also consistent with the experimental observation of the strong $T$ dependence of $\mu_{FE}$ in MIS-MIS and MS-MS FETs (see **Fig. 5 (c)**).

In conclusion, we designed and fabricated the innovative device architecture to serve a new platform to investigate the 2D MS and MIS contacts in both symmetric and asymmetric conditions. We confirmed the improvement of the contact condition and consequently the enhancement of the device performance of MoS$_2$ FETs by exploiting the MIS contacts decorated with the monolayer

*h*-BN. Moreover, we revealed the novel diode-like selective enhancement of the carrier transport through the MIS contact. The MIS contact can significantly reduce the contact resistance and boost the electron transport from the semiconductor to the metal, but barely affect the electron transport oppositely. With the concept of the carrier collection barrier, we revealed the underlying physics of the selective enhancement of the carrier transport through the MIS contact. We also interpreted the negative SBH obtained from the *T*-dependent measurement using the carrier collection barrier, in a comparison with the positive SBH described by the carrier injection barrier. Our work has advanced the fundamental understanding of 2D MIS contacts, and demonstrated the critical role of the collection barrier in the carrier transport of 2D nanoelectronic devices.

## Methods

**Material synthesis and characterization.** The monolayer $MoS_2$ was synthesized using a customized two-zone chemical vapor deposition (CVD) system [53]. Specifically, ammonium heptamolybdate (AHM) was used as a water-soluble Mo precursor, and NaOH was used as a water-soluble promoter. Their mixed solution was spin-coated on a $SiO_2$/Si growth substrate. The reaction between Mo and Na produced $Na_2MoO_4$ compounds and then become $MoS_2$ after S vapor injection. We optimized the annealing time to control the size of the isolated monolayer triangular domains. After synthesis, the monolayer $MoS_2$ flakes were wet-transferred onto a $SiO_2$/Si device substrate. The large-area CVD-grown monolayer *h*-BN was purchased from 6Carbon Technology [54]. The Raman spectroscopy was performed by Renishaw inVia Raman microscope. The AFM microscopy was performed by Bruker Dimension Icon with ScanAsyst.

**Device fabrication and measurement.** Both the FET and TLM devices were fabricated using electron-beam lithography (EBL) and evaporation with a Ti/Au (10 nm/100 nm) metal layer. Specifically, three types of global-back-gate transistors, including an MS-MS FET, an MIS-MIS FET, and an MS-MIS (or MIS-MS) FET, were fabricated on a single monolayer $MoS_2$ triangular domain with the identical channel length and width (see **Fig. 1** and **Fig. S1** in **Supporting Information**). To ensure a consistent channel width for the comparison of all the types of the transistors, here we took the average value of the source and drain contact lengths as the channel width for the trapezoidal geometry of the channel. All devices were fabricated on n-type Si substrates (0.001-0.005 $\Omega$cm) which have 285 and 90 nm $SiO_2$ layers. The electrical measurements were performed in a vacuum-chamber probe station (MSTECH M5VC) with a semiconductor parameter analyzer (Keysight B1500A), and the temperature varied from 203 to 303 K.

## Data availability

The data that support the findings of this study are available from the corresponding author on reasonable request.

[12] D. S. Schulman, A. J. Arnold, and S. Das, "Contact engineering for 2D materials and devices", *Chem. Soc. Rev.*, 47, 3037-3058, 2018.

[13] Y. Zhao, K. Xu, F. Pan, C. Zhou, F. Zhou, and Y. Chai, "Doping contact and interface engineering of two-dimensional layered transition metal dichalcogenides transistors", *Adv. Funct. Mater.*, 27, 1603484, 2016.

[14] A. Allain, J. Kang, K. Banerjee, and A. Kis, "Electrical contacts to two-dimensional semiconductors", *Nat. Mater.*, 14, 1195-1205, 2015.

[15] D. Jena, K. Banerjee, and G. H. Xing, "2D crystal semiconductors: Intimate contacts", *Nat. Mater.*, 13, 1076-1078, 2014.

[16] A. D. Franklin, "Nanomaterials in transistors: From high-performance to thin-film applications", *Science*, 349, 6249, 2015.

[17] M. S. Choi, D. Qu, D. Lee, X. Liu, K. Watanabe, T. Taniguchi, and W. J. Yoo, "Lateral $MoS_2$ p-n junction formed by chemical doping for use in high-performance optoelectronics", *ACS Nano*, 8, 9, 9332-9340, 2014.

[18] H. Li, D. Lee, D. Qu, X. Liu, J. Ryu, A. Seabaugh, and W. J. Yoo, "Ultimate thin vertical p-n junction composed of two-dimensional layered molybdenum disulfide", *Nat. Commun.*, 6, 6564, 2015.

[19] H. Fang, M. Tosun, G. Seol, T.C. Chang, K. Takei, J. Guo, and A. Javey, "Degenerate n-doping of few-layer transition metal dichalcogenides by potassium", *Nano Lett.*, 13, 5, 1991-1995, 2013.

[20] D. Kiriya, M. Tosun, P. Zhao, J. S. Kang, and A. Javey, "Air-stable surface charge transfer doping of $MoS_2$ by benzyl viologen", *J. Am. Chem. Soc.*, 136, 22, 7853-7856, 2014.

[21] A. Kutana, E. S. Penev, and B. I. Yakobson, "Engineering electronic properties of layered
16

## Acknowledgments

This work was partially supported by the National Science Foundation (NSF) under Award ECCS-1944095, the New York State Energy Research and Development Authority (NYSERDA) under Award 138126, and the New York State Center of Excellence in Materials Informatics (CMI) under Award C160186. The authors acknowledge support from the Vice President for Research and Economic Development (VPRED) and Quantum Science and Technology (QST) at the University at Buffalo. A.C. acknowledges support from the Presidential Fellowship Program at the University at Buffalo.

## Author contributions

F.Y. and H.L. conceived and supervised the project. J.L., C.C., and F.Y. synthesized the $MoS_2$ samples. H.N.J. and R.D. fabricated the 2D FET and TLM devices. H.N.J., M.L., and S.S. performed the electrical characterizations. S.W. performed the material characterizations. A.C., Y.G., R.W., and J.M.L. participated in the sample preparation and device characterization. X.L. and C.Y. participated in the data analysis.

## Ethics declarations

The authors declare no competing interests.

# Figures

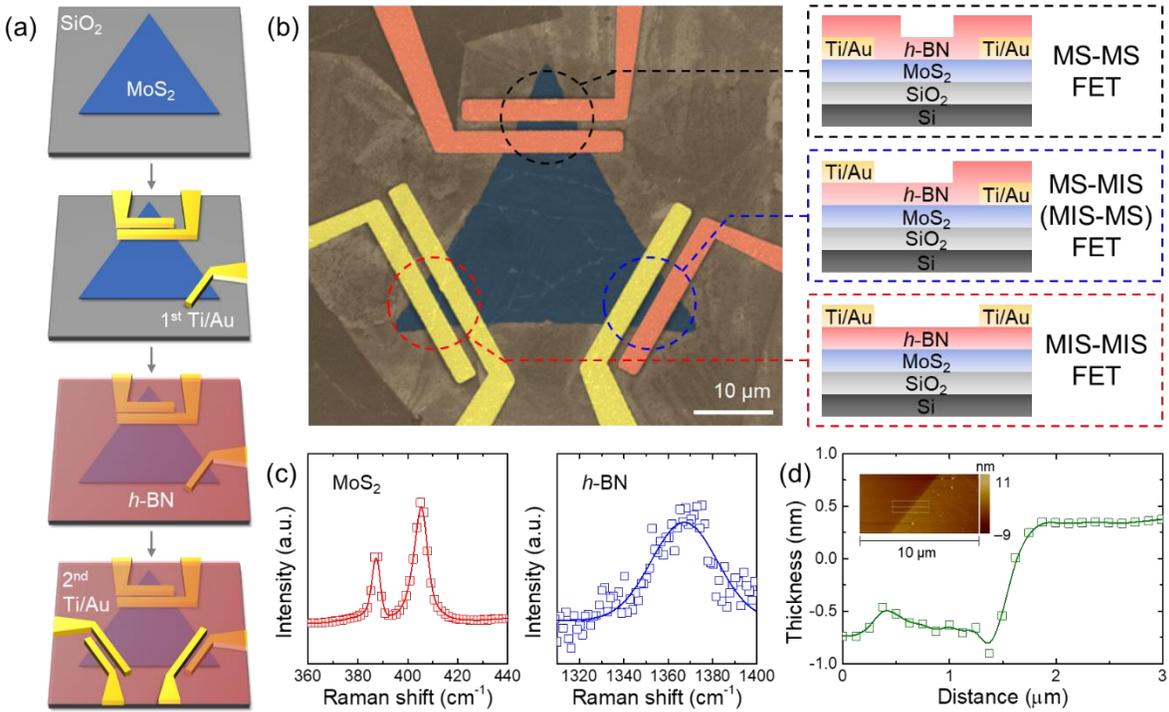

**Fig. 1 Multiple MoS₂ FETs with MS and MIS contacts on a single monolayer MoS₂ triangular domain.** (a) Schematic illustration of the fabrication process. (b) False-colored SEM image of the MS-MS, MS-MIS (MIS-MS), and MIS-MIS FETs on a single MoS$_2$ triangular domain as well as the corresponding cross-section schematics of the device structure. (c) Raman spectrum of the monolayer MoS$_2$ and $h$-BN. (d) AFM image (inset) and scan profile of the monolayer MoS$_2$.

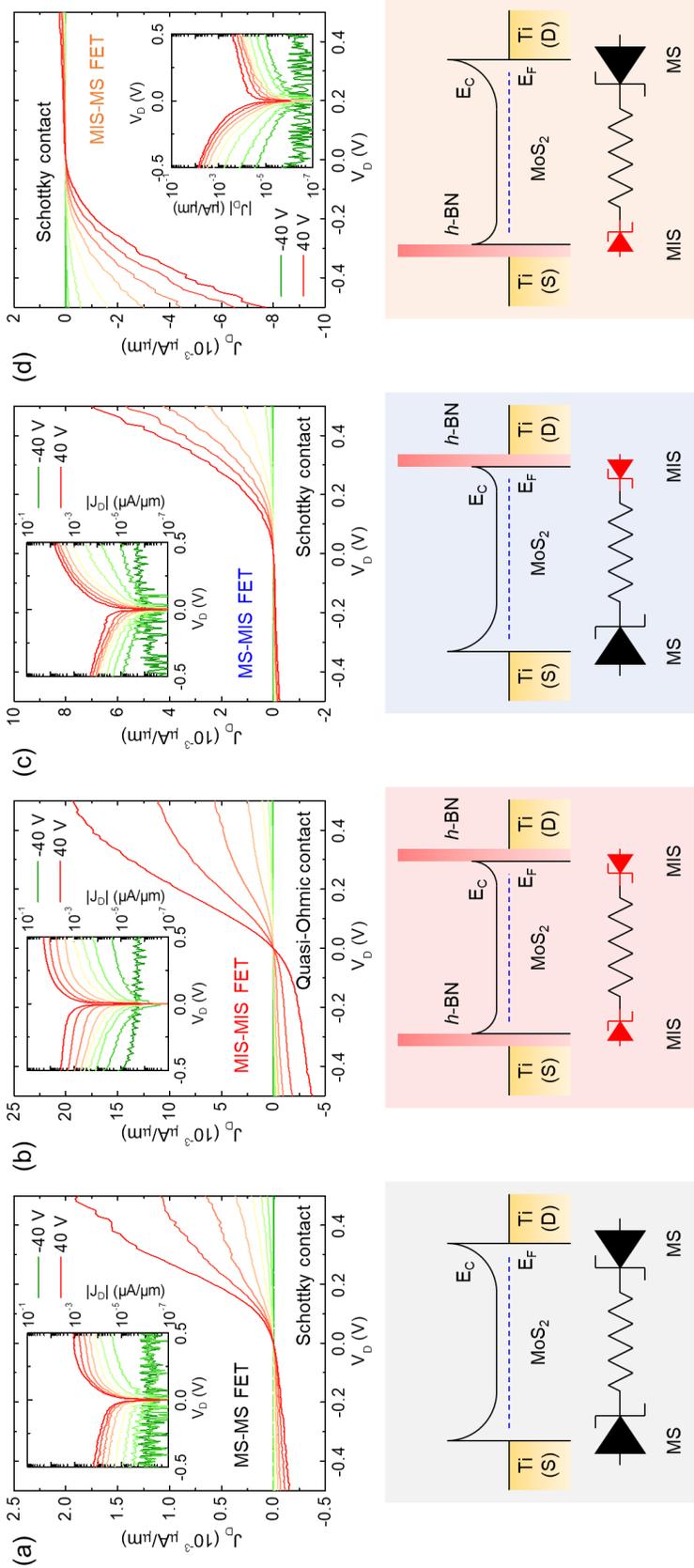

**Fig. 2 Comparison of output characteristics of MoS$_2$ FETs with MS and MIS contacts.** (a-d) Top panel: Room-temperature $J_D$-$V_D$ characteristics in linear scale as well as logarithmic scale (inset) for the MS-MS, MIS-MIS, MS-MIS, and MIS-MS MoS$_2$ FETs, respectively. $V_G$ varies from –40 V (green) to 40 V (red) with a step of 10 V. Bottom panel: The corresponding energy band structures along the channel and the equivalent electronic circuits for all the types of the FETs with the MS and MIS contacts. For the ease of illustration and comparison, the effect of the applied $V_D$ is not included. Here $E_C$ and $E_F$ denote the conduction band edge and the Fermi energy level, respectively.

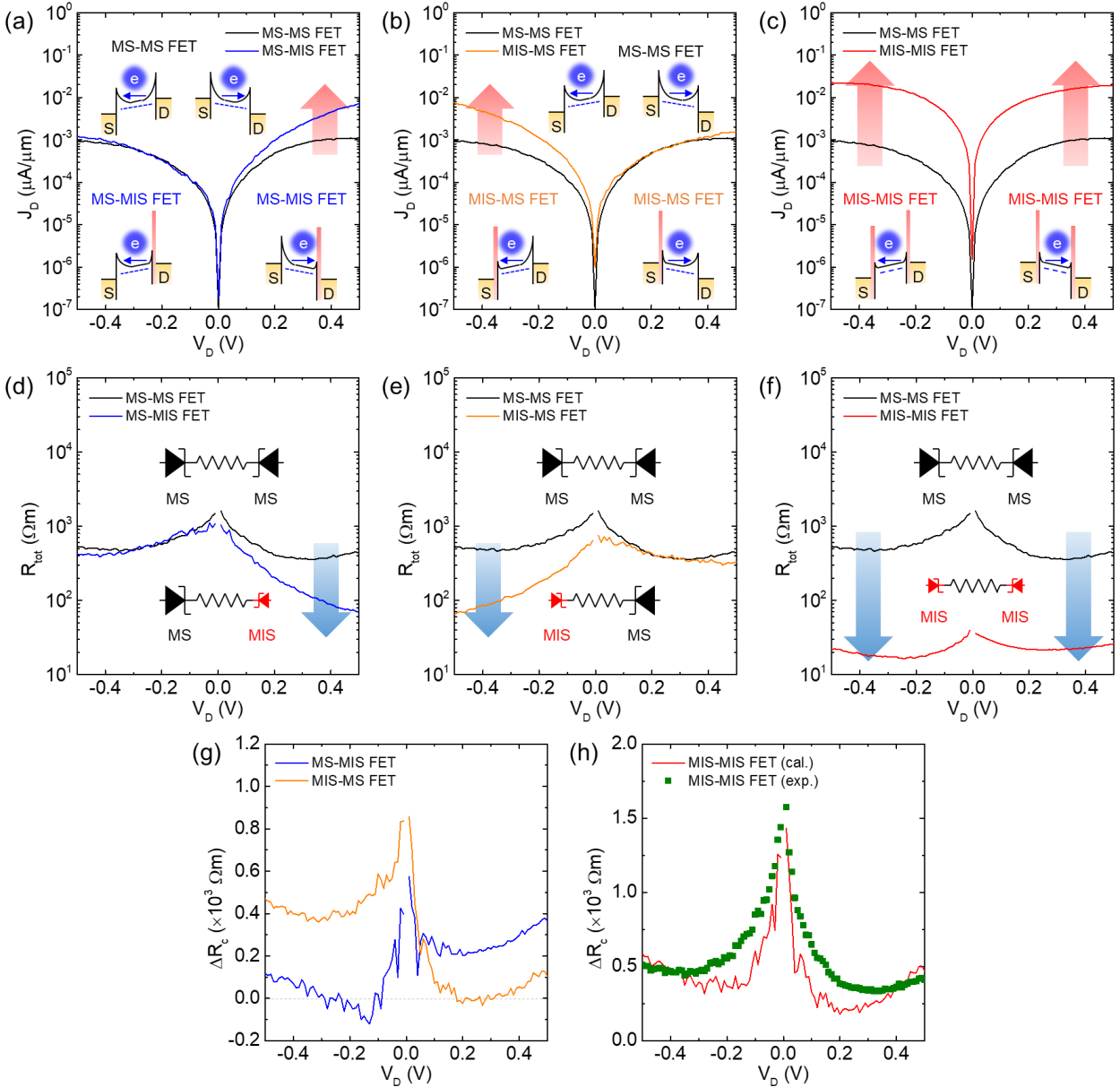

**Fig. 3 Diode-like selective enhancement of carrier transport through MIS contact.** (a-c) Comparison of the revised $J_D$-$V_D$ characteristics ($V_G$ = 40 V) by including the asymmetric factor for the MS-MS, MS-MIS, MIS-MS, and MIS-MIS FETs, respectively. The MS-MS FET is used as a control sample. (d-f) Comparison of the corresponding $R_{tot}$ for the MS-MS, MS-MIS, MIS-MS, and MIS-MIS FETs, respectively. (g, h) The calculated $\Delta R_c$ as a function of $V_D$ for the MS-MIS, MIS-MS, and MIS-MIS FETs, in a comparison with the experimental value obtained from the MIS-MIS FET.

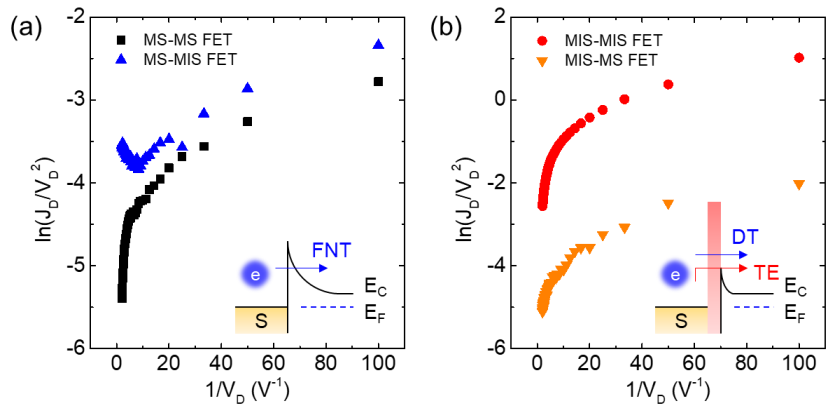

**Fig. 4 Impact of MS and MIS drain contacts on the carrier injection at the source.** (a, b) Comparison of $\ln(J_D/V_D^2)$ versus $1/V_D$ curves at $V_G = 40$ V for the MS-source FETs (including the MS-MS and MS-MIS FETs) and the MIS-source FETs (including the MIS-MIS and MIS-MS FETs), respectively. Inset: Energy band diagram illustrates the FNT dominating in the MS-MIS FET, in a comparison with the DT and TE dominating in the MS-MS, MIS-MIS, and MIS-MS FETs.

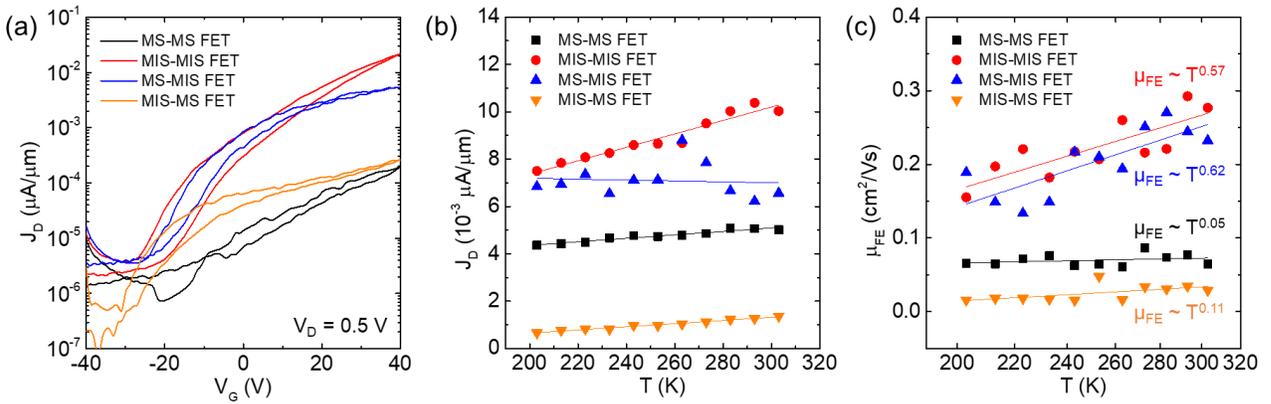

**Fig. 5 Comparison of transfer characteristics of MoS$_2$ FETs with MS and MIS contacts.** (a) Room-temperature $J_D$-$V_G$ characteristics in logarithmic scale for the MS-MS, MIS-MIS, MS-MIS, and MIS-MS MoS$_2$ FETs at $V_D = 0.5$ V. (b) The maximum $J_D$ obtained at $V_G = 40$ V as a function of $T$ for all the types of the FETs. (c) The maximum $\mu_{FE}$ as a function of $T$ in logarithmic scale for all the types of the FETs.

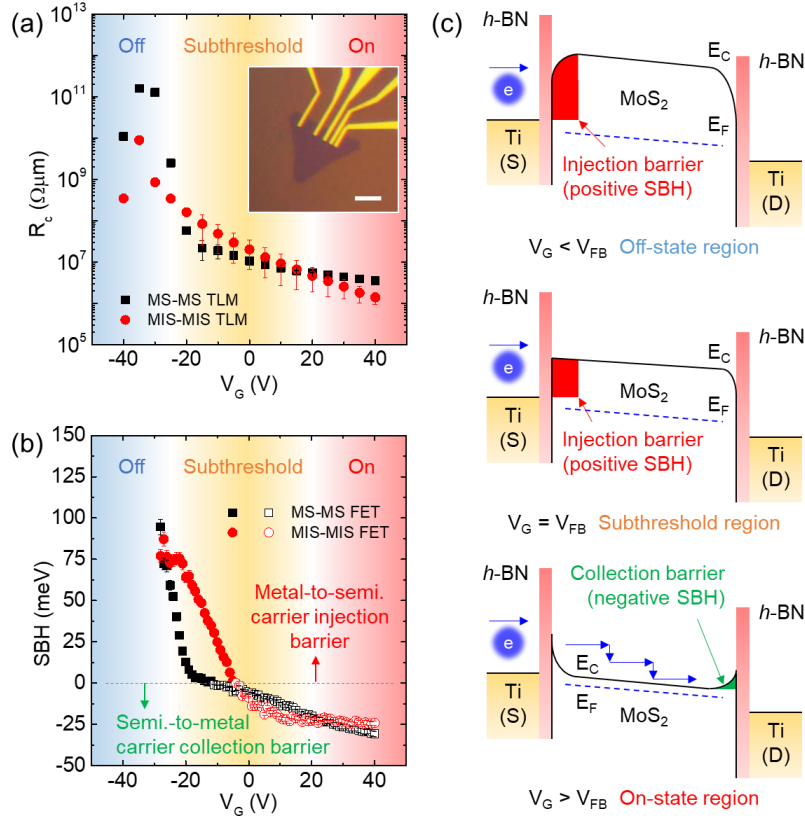

**Fig. 6 Comparison of contact resistance and barrier height with MS and MIS contacts.** (a) Room-temperature $R_c$ as a function of $V_G$ for the MS-MS and MIS-MIS TLM devices. The blue, yellow and red backgrounds indicate the off-state, subthreshold, and on-state regions, respectively. Inset: Optical microscope image of an MIS-MIS TLM device. Scale bar: 10 μm. (b) SBH as a function of $V_G$ for the MS-MS and MIS-MIS FETs. The positive and negative SBHs indicate metal-to-semiconductor carrier injection barrier height (solid symbol) and semiconductor-to-metal carrier collection barrier height (hollow symbol), respectively. (c) The corresponding energy band diagram along the channel for the MIS-MIS FET at a positive $V_D$ and various $V_G$ conditions ($V_G < V_{FB}$, $V_G = V_{FB}$, and $V_G > V_{FB}$, where $V_{FB}$ is the flat-band gate voltage). The predominant injection barrier at the off-state and subthreshold regions is highlighted in red, and the collection barrier at the on-state region is highlighted in green.

# Supporting Information

# Diode-like Selective Enhancement of Carrier Transport through Metal-Semiconductor Interface Decorated by Monolayer Boron Nitride

Hemendra Nath Jaiswal,[1] Maomao Liu,[1] Simran Shahi,[1] Sichen Wei,[2] Jihea Lee,[2] Anindita Chakravarty,[1] Yutong Guo,[1] Ruiqiang Wang,[1] Jung Mu Lee,[1] Chaoran Chang,[2] Yu Fu,[2] Ripudaman Dixit,[1] Xiaochi Liu,[3] Cheng Yang,[4] Fei Yao,[2*] and Huamin Li[1*]

[1] *Department of Electrical Engineering, University at Buffalo, The State University of New York, Buffalo, New York 14260, US*

[2] *Department of Materials Design and Innovation, University at Buffalo, The State University of New York, Buffalo, New York 14260, US*

[3] *School of Physics and Electronics, Central South University, Changsha, 410083, China*

[4] *School of Physics and Electronics, Shandong Normal University, Jinan 250014, China*

[*] Author to whom correspondence should be addressed. Electronic address: feiyao@buffalo.edu and huaminli@buffalo.edu

# Supporting Discussion

## 1. Asymmetric IV characteristics

It is possible that the contact area may not be the sole source for the asymmetric IV behavior as shown in Fig. 2, but in this work it should've played a dominating role compared to other factors, for example, the local defects, oxidation, and drain bias as mentioned in a recent work [S1]. This conclusion is evidenced by our experimental results, i.e. the comparison of the MS-MS and MIS-MIS devices and the comparison of the MS-MIS and MIS-MS devices. Specifically, the first comparison shows an identical rectification effect (same asymmetric coefficient as a function of $V_D$) under two totally different contact conditions (one with the BN decoration layer and another one without it). This can exclude the possible effects from the local defects or oxidation at the contact interface. The second comparison is performed on the same device but the source and drain are switched for each measurement. Again, the identical rectification effect is obtained. This can exclude the possible effects from the applied drain voltage. Overall, our experimental results indicate that the asymmetric IV behavior is predominately induced by a trapezoidal geometry of the channel and thus the different contact areas at the source and drain.

## 2. Drain voltage effect on energy band diagram

For ease of the schematic illustration, the $V_D$ effect in the band diagram is not included in Fig. 2. We understand that an applied positive $V_D$ will lower the drain side and generate a slope along the channel. Together with an image force barrier lowering effect, a minor asymmetry between the source and drain contacts can be induced in principle. Especially, a lowered barrier at the drain end is expected, which can enhance the electron transport from the $MoS_2$ channel to the drain electrode. However, we don't think this physical model is needed to be considered in our case because we are using a much smaller $V_D$ (up to +/− 0.5 V for a channel length of 2 μm).

In general, the asymmetric band structure is considered only for the case of the asymmetric metallization, i.e., a high workfunction metal contact at one end in contrast with a low workfunction metal contact at another end [S2]. For the case of the symmetric metallization, a symmetric energy band structure between the source and drain is considered for most of the 2D devices, even with an applied $V_D$ voltage [S3]. Especially, scanning photocurrent microscopy has experimentally demonstrated that a nearly symmetric band structure between the source and drain contacts under a small $V_D$ [S4]. A clear barrier at the drain end indicates a resistance to prevent the carrier transport from the semiconductor to the drain contact. Although the charge carriers in these studies were excited by photon illumination, the conventional charge carriers induced by thermal generation should also face the same collection barrier.

To precisely present the impact of the applied $V_D$ and its fractional drop on the source and drain contacts as well as the channel, $E_F$ of the MoS$_2$ channel cannot be aligned with $E_F$ of the contacts. Therefore, we included the $V_D$ effect (both the positive and negative $V_D$) as well as the $E_F$ shift among the source, channel, and drain in the energy band diagrams, as shown in Fig. 3 (a-c).

### 3. Charge carrier mobility in MoS$_2$

The monolayer MoS$_2$ triangular domain was synthesized in the lab using a customized two-zone CVD furnace. Due to the imperfection of the synthesis condition, the charge carrier mobility of the monolayer MoS$_2$ is lower compared to other samples, either from mechanical exfoliation or from commercialized CVD products. This is exactly the reason we innovatively design and fabricate a platform on a single MoS$_2$ domain so all the devices share the same quality of the channel material for comparison. In this way, we can exclude all the other factors and concentrate on the contact condition which is the only variable in the comparison. The synthesized MoS$_2$ quality can be further improved by optimizing the synthesis parameters, but our conclusion from this work won't change.

## Supporting Figures

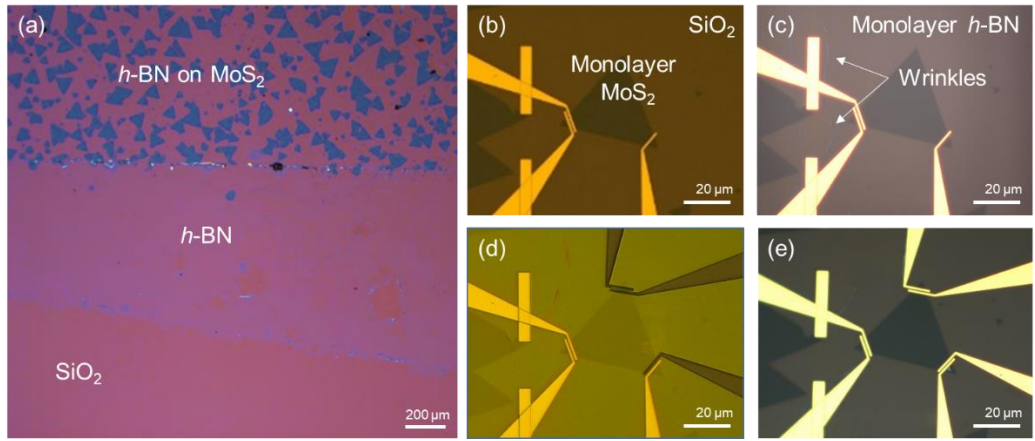

**Fig. S1** (a) Optical microscope image of an $h$-BN/MoS$_2$ heterostructure on SiO$_2$/Si substrate. The monolayer MoS$_2$ and monolayer $h$-BN are wet-transferred on the substrate sequentially. (b-e) Optical microscope images show the process to fabricate the MS-MS, MS-MIS (MIS-MS), and MIS-MIS FETs on a single monolayer MoS$_2$ triangular domain.

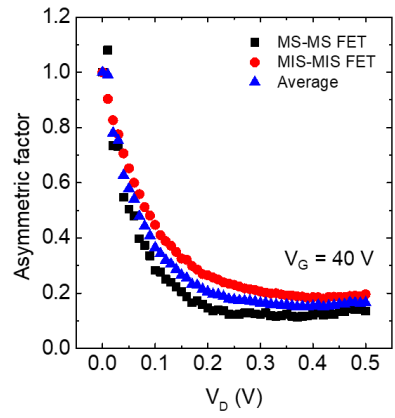

**Fig. S2** Asymmetric factor versus $V_D$ for the MS-MS and MIS-MIS FETs at $V_G = 40$ V and the average value. The asymmetric factor is defined as the ratio of the magnitude of the current density at the positive $V_D$ to that at the negative $V_D$.

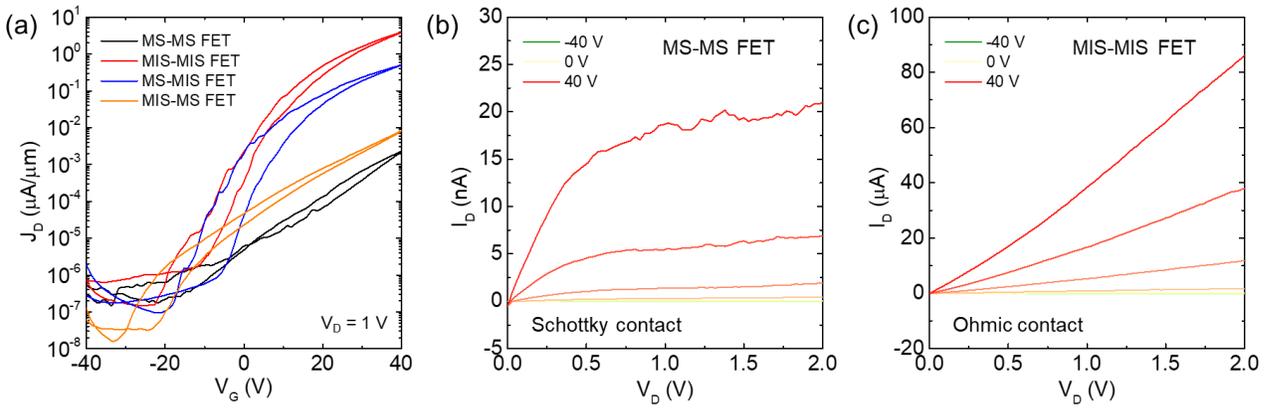

**Fig. S3** Comparison of transistor performance of MoS$_2$ FETs with MS and MIS contacts, in addition to the ones shown in Fig. 5 (a). (a) Room-temperature $J_D$-$V_G$ characteristics at $V_D$ = 1 V are plotted in logarithmic scale for the MS-MS, MIS-MIS, MS-MIS, and MIS-MS MoS$_2$ FETs. The device overall shows a better performance in terms of on-off ratio, on-current density, and subthreshold swing compared to the one in Fig. 5 (a), due to a thinner SiO$_2$ (90 nm) used for the back gating and a higher $V_D$ (1 V). (b, c) Room-temperature $I_D$-$V_D$ output characteristics of the MS-MS and MIS-MIS FETs, respectively. $V_G$ varies from –40 V (green) to 40 V (red) with a step of 10 V. Both the devices show the strong enhancement of the carrier transport and the Schottky-to-Ohmic contact improvement by exploiting the MIS contact, and demonstrate the reliability and repeatability of the MIS decoration technique.

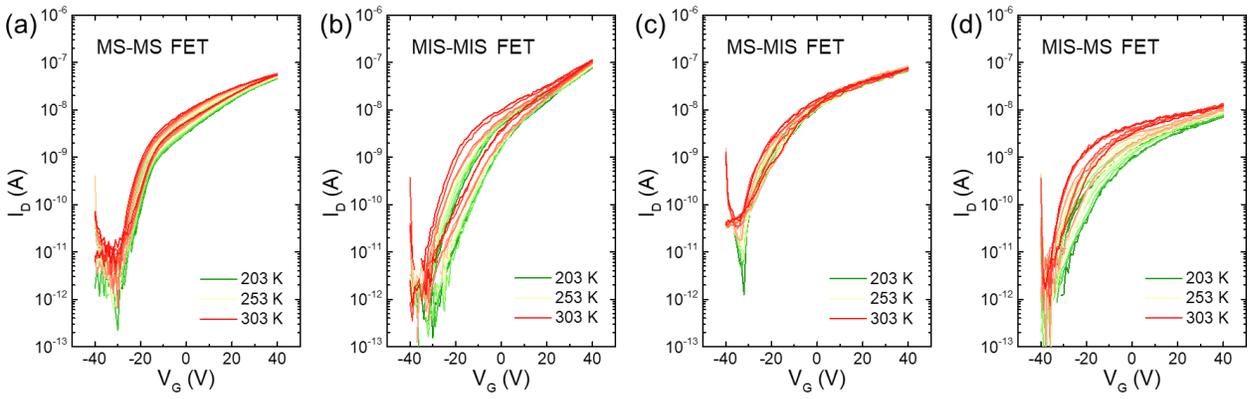

**Fig. S4** $T$-dependent $I_D$-$V_G$ characteristics at $V_D$ = 0.5 V for the MS-MS, MIS-MIS, MS-MIS, and MIS-MS FETs. $T$ ranges from 203 K (green) to 303 K (red) with a step of 10 K.

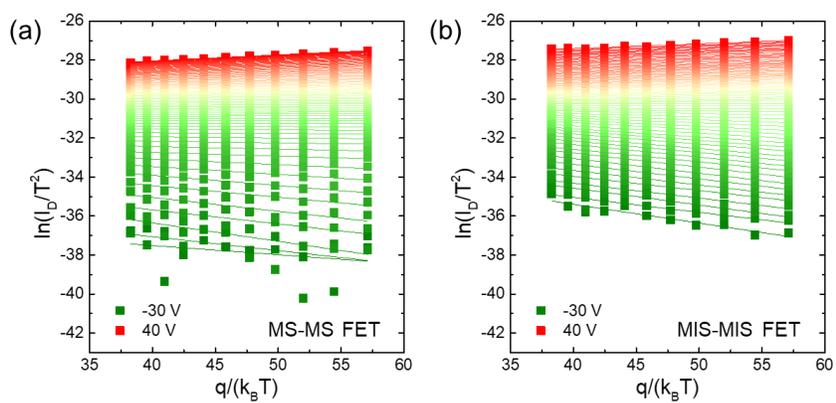

**Fig. S5** ln($I_D/T^2$) versus $q/(k_BT)$ plots for (a) the MS-MS FET and (b) the MIS-MIS FET with the $T$ ranging from 203 to 303 K. The value of the SBH is extracted from the linear fit as a function of $V_G$ ranging from –30 V (green) to 40 V (red).